# Navigating Compiler Errors with AI Assistance - A Study of GPT Hints in an Introductory Programming Course

Maciej Pankiewicz
Warsaw University of Life Sciences
Warsaw, Poland
maciej_pankiewicz@sggw.edu.pl

Ryan S. Baker
University of Pennsylvania
Philadelphia, USA
ryanshaunbaker@gmail.com

## ABSTRACT

We examined the efficacy of AI-assisted learning in an introductory programming course at the university level by using a GPT-4 model to generate personalized hints for compiler errors within a platform for automated assessment of programming assignments. The control group had no access to GPT hints. In the experimental condition GPT hints were provided when a compiler error was detected, for the first half of the problems in each module. For the latter half of the module, hints were disabled. Students highly rated the usefulness of GPT hints. In affect surveys, the experimental group reported significantly higher levels of focus and lower levels of *confrustion* (confusion and frustration) than the control group. For the six most commonly occurring error types we observed mixed results in terms of performance when access to GPT hints was enabled for the experimental group. However, in the absence of GPT hints, the experimental group's performance surpassed the control group for five out of the six error types.

## 1 Introduction

Hints are a part of learning environments aimed at helping learners understand a given task or a set of concepts better [23]. There are many examples of systems from the area of intelligent tutoring systems [3, 4, 13] and other online educational environments [7, 9], including programming [16, 20] that use hints to support self-paced learning. The impact of this kind of intervention on learning has been studied in previous research and mixed results have been observed. Positive effects on learning outcomes and engagement have been identified [1, 25] but also negative effects such as hint abuse have been reported [2].

In programming education, in an introductory course for computer science students, where compiled languages such as Java, C++ or C# are usually used, we would expect hints to address several distinct areas: resolving a syntax error (when the code doesn't compile), a runtime error (when the code compiles, but it throws an exception during program execution), or an error in execution logic (when the program executes, but it does not generate expected outcome). Each of these areas separately presents different challenges to hint generation.

In this study we focus on automating generation of hints for novices to help them in resolving compiler errors. In general, design of compiler messages impedes learning for beginners [24], while students' ability to resolve compiler errors is one of the key factors in the early stages of learning to program [5, 6, 10, 11]. There are several factors that make it challenging to automate hint generation for compiler errors to support novices at their initial steps in introductory programming courses: 1) the number of messages given for different syntax errors is large (exceeding 1400 in the case of C#), 2) different compilers (even for the same programming language) may report different messages for the same code (or code structure), 3) the root for each of these errors is in (usually) unique code created by a student.

Large language models (LLMs) have potential to address these issues efficiently: they are trained on a large corpus of programming data [8], are context aware [21] and preliminary results show that hints automatically generated by LLMs may be effective for programming education [18]. Therefore, to specifically address the issue of hint generation for compiler errors and their impact on beginners taking an introductory programming course, we conducted an experimental study using a platform for automated assessment of programming assignments, and tested the impact of hints generated dynamically through an API using the GPT-4 engine.

## 2 Methods

Data for this study was collected during the first 3 months (2nd October – 31st December) of the 4-month-long winter semester 2023/2024. Participants were computer science students taking a mandatory first-semester "Introduction to Programming" course at a large university in Poland. The programming language for the course was C#. Consent was obtained from students prior to joining this study.

A total of 259 students (29% female) consented to participate with N=97 (44% female) being novices in programming, as determined by a pre-course questionnaire. This questionnaire asked students about their programming knowledge on a scale from 1 (no experience) to 5 (extensive experience), asking about fundamental concepts such as types, variables, conditional statement, recursion, loops, and arrays. The study defined novices as those who rated themselves with a 1 or 2 on this scale, excluding participants who rated their experience level as 3 to 5. The accuracy of students' self-assessed programming expertise was investigated through a pre-test aimed at objective evaluation of their knowledge, administered directly after the questionnaire. The difference in pre-test scores between groups of novices (Mdn=24%) and more experienced students (Mdn=71%) was statistically significant (W=2047, p<0.001) for a non-parametric Mann-Whitney U test.

We utilized the runcodeapp.com platform for automated assessment of programming assignments, sharing a total of 146 programming problems with students [19]. The following topics were covered: types and variables (33 problems), conditional statement (25), recursion (28), loops (17), and loops with arrays (43).

Tasks were categorized into 14 modules each containing from 5 to 14 problems. Task difficulty varied, with the easiest tasks requiring to multiply two integers, to much more difficult tasks, such as extracting and alphabetically sorting words from a given text. The platform's assessment process involved compiling and testing the submitted code, followed by providing students with feedback that included: compiler messages (line, error message and id), results for each unit test executed on the compiled code (with input values and expected outcome for each unit test), and overall score (the percentage of unit tests that ended successfully: 0-100%). Usage of the platform was not mandatory and performance did not count towards the final grade.

In this study we specifically focus on beginners (we also refer to them as novices). Participants with this level of experience were assigned into two groups: a control group (N=48), and an experimental group (N=49). No significant difference was identified between the median pre-test score of the experimental group (Mdn=21%) and the control group (Mdn=24%) for a non-parametric Mann-Whitney U test (W=1181.5. p=0.971). In the experimental condition, GPT hints (GPT-4 model: gpt-4-0613) were generated in Polish directly after the code submission, when a compiler error was detected, and provided as an additional platform feature for the first half of the problems in each module. In the case of a submission containing multiple compiler errors, a GPT hint was generated for the first error identified by the compiler.

Hints were created via a request to the OpenAI API using the approach to prompt generation proposed in [18] and consisted of the following parts: 1) general instructions for an assistant (in English), 2) assignment text (in Polish, as it is defined on the platform), 3) student code, 4) results of the code evaluation (in English). The prompt was extended by examples of an ideal response containing three elements (in English): an explanation of the error (example from one of generated hints: "The compiler message `;  expected` means that a semicolon is missing in the line `int a = 2*b`"), a solution strategy (example: "To fix this error, you need to add a semicolon `;` at the end of the line `int a = 2*b`"), and an educational element regarding the underlying concept (example: "Remember that in C#, every statement must be terminated with a semicolon"). The assistant was instructed to generate the hint in Polish. The hint was generated immediately after the submission (it took approximately 15-20 seconds to generate it) and was presented as a pop-up. Students could close it and return back to work. After closing the pop-up, students could again view this hint by clicking a button below the code editor. The hint was generated once per submission. No limit was imposed on the number of submissions. For the latter half of the module, no hints were offered to evaluate the impact of support received from GPT on students' subsequent platform performance. In total, GPT hints were generated for 71 out of 143 tasks.

While solving tasks on the platform students self-reported their affective states through a dynamic HTML element presented directly after they received submission results with the question: "Choose the option that best describes how you feel at the moment" with the following response options: *Focused*, *Anxious*, *Confused*, *Frustrated*, *Bored*, *Other* (in this order) and appropriate emoticons to visually highlight each of the available responses. This set of affective states was chosen based on their importance for learning [12]. To avoid irritation that could arise from being required to fill out the affect survey too frequently, it was not presented after every submission, but randomly, with a probability of 1/6. In previous research, it has been found that unresolved confusion and frustration negatively impact learning outcomes [14, 15]. In subsequent research the impact of these two affective states has been analyzed under an umbrella of *confrustion* [17, 22] and we use this combined affective state in our analysis.

To evaluate the usefulness of GPT-generated hints, a dynamic HTML element was presented to students in the experimental condition with a prompt asking to rate the usefulness of received hints on a 5-point Likert scale ranging from '1 – Not useful at all' to '5 – Extremely useful'. After every fifth GPT hint the student received, the student was given this survey.

## 3 Results

Out of 97 novices, 61 students submitted at least one solution on the platform during the study period, generating in total 14,830 submissions (7,640 – control, 7,190 – experimental: 3,487 submissions to tasks with GPT hints enabled, and 3,703 for the remaining tasks). In the experimental condition 1,222 GPT hints were generated for submissions containing a compiler error on tasks with GPT hints enabled. Erroneous code submitted during the study contained 110 different syntax errors.

In the following sections, we analyze the reception of the GPT hint feature by students, and examine its impact on affect immediately following the submission of code containing a compiler error. Additionally, we assess performance, looking at the fraction of compiler errors in tasks where GPT hints were either enabled or disabled for the experimental group. To provide insight into the effect of the intervention and the ability to resolve errors after disabling GPT hints, our analysis of affect state and student performance focuses on submissions that contained exactly one compiler error. Our aim is to provide insights into both the perceived and actual effectiveness of the GPT-hint feature in the context of computer science education.

### 3.1 Hint usefulness

To evaluate the perceived usefulness of the GPT generated hints, a total of 213 collected responses from N=25 students in the experimental group were analyzed. The control group did not have access to the GPT hints, therefore no responses were recorded from this group. Figure 1 displays the histogram of the frequency of students according to the median rating they gave to received hints, with the x-axis representing the median of the hint rating for a user.

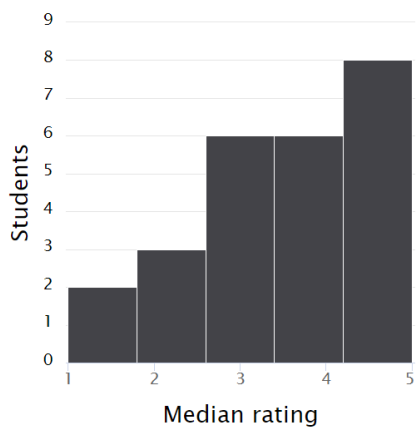

**Figure 1: Distribution of students (N=25) according to the median rating for received hints**

The majority of students (56%) generally rated the hints as *very* or *extremely* useful – this shows a generally positive reception of the GPT hints feature by students. However, 20% of students generally rated hints as *not useful at all* or *slightly* useful.

### 3.2 Affective states

To understand the impact of the introduced feature on student affect we analyzed a total of 1401 survey responses (experimental: 754, control: 647) from 54 students (experimental: 27, control: 27) collected during the period of study with *focused* being the most frequently reported state in both conditions. Out of these responses, 210 have been reported while solving tasks with GPT hints enabled, directly after receiving a hint for a submission with one compiler error.

To compare differences in the frequencies of the affective states reported by each student after receiving a hint for an error, non-parametric Mann-Whitney U tests were calculated, due violation of normality assumptions. Due to multiple comparisons, the Benjamini-Hochberg alpha correction was used. Two affective states showed marginally significant differences between conditions. For *focused*, students in the experimental group (N=15) reported marginally higher rates (Mdn=0.9) than those in the control group (Mdn=0.25, N=13), (W=153, adjusted α=0.0125, p=0.0152, for a non-parametric Mann-Whitney U test). For *confrustion* (described in section 2), students in the experimental group reported lower rates (Mdn=0) than those in the control group (Mdn=0.0167), (W=48, adjusted α=0.025, p=0.0157, for a non-parametric Mann-Whitney U test). *Boredom* and *anxiety*, however, did not show significant differences between conditions. For *boredom*, the experimental group reported a median that was not significantly different (Mdn=0) from the control group (Mdn=0), (W=96.5, adjusted α=0.05, p=0.958, for a non-parametric Mann-Whitney U test). The same was true for *anxiety* (Exp. Mdn=0 vs. Contr. Mdn=0.077), (W=81.5, adjusted α=0.075, p=0.444, for a non-parametric Mann-Whitney U test).

### 3.3 Compiler errors: student performance

To compare the impact of GPT hints on student performance between 1) when these hints were available and 2) after they were disabled, we examined the first five attempts made for all tasks available on the platform by each student (each student-task pair), focusing on the presence of compiler errors in these submissions.

The study was conducted under conditions differentiated by the availability of GPT-generated hints. This is visualized in our charts, where the line marked in gray color (Phase 1) represents submissions for tasks where GPT hints were enabled for the experimental group (first half of the problems listed for each of the modules). In contrast, the line in black color (Phase 2) provides insight into the submissions for more difficult tasks where GPT hints were disabled (second half of the problems listed for each of the modules). As a reminder: for both sets of tasks the control group did not have access to GPT hints.

This approach allowed us to explore the impact of AI-generated hints on the frequency of compiler errors, offering insights into the effectiveness of AI tools in enhancing error-resolving competency.

The presented charts illustrate the fraction of code submissions in which a particular compiler error was detected across successive attempts (lower values mean better performance). Each data point represents the total percentage of submissions with the identified compilation error, cumulatively calculated up to that specific attempt. In these calculations, the first five attempts on each task available on the platform were included.

Due to a large number of different errors identified in student submissions, we constrain our analysis only to the compiler errors most prevalent in our dataset with the following IDs: CS1002 (8%), CS0103 (5%), CS0266 (4%), CS1525 (3%), CS0161 (3%), CS1026 (2%).

*3.3.1 CS1002 "; expected".* This compiler error occurs when a semicolon, which is required at the end of a statement, is missing. For this error we observed that in scenarios where GPT hints were accessible (Phase 1), the performance of the experimental group was slightly lower than that of the control group (Figure 2). However, in the set of more complex tasks where GPT hints were withdrawn from the experimental group (Phase 2), the performance of the experimental group only saw a minor decline and was distinctly better than in the control group. For the control group, we observed a significant drop in performance on the more complex tasks.

These results indicate that initial exposure to AI-assisted error resolving, even when later removed, may have potential to impart lasting benefits in tackling more complex programming assignments.

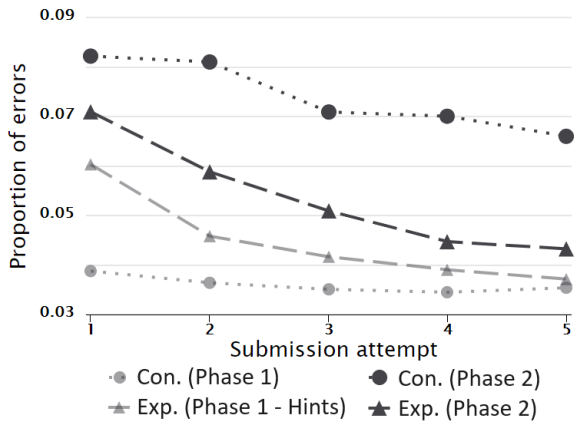

**Figure 2: Error CS1002 "; expected". Proportion of Errors Across First 5 Submission Attempts (cumulative).**

*3.3.2 CS0103 "The name '{0}' does not exist in the current context".* Similar outcomes are seen for the CS0103 compiler error, occurring when a variable or method name has not been declared or is not accessible in the scope where it is being used.

In this case, for the set of tasks where GPT hints were activated (Phase 1), the experimental group displayed lower performance than the control group (Figure 3).

However, in the case of more advanced tasks (Phase 2), while the control group was slightly less successful at fixing this error, the performance of the experimental group improved, surpassing that of the control group as in CS1002.

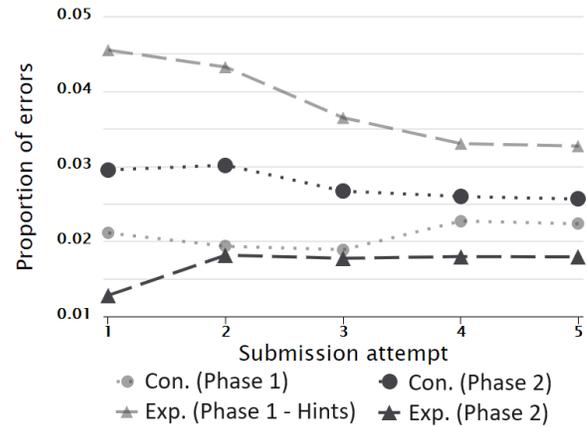

**Figure 3: Error CS0103 "The name '{0}' does not exist in the current context". Proportion of Errors Across First 5 Submission Attempts (cumulative).**

*3.3.3 CS1525 "Invalid expression term '{0}'".* This error is often a result of typos or syntax misunderstandings and typically occurs when there's an invalid term in an expression, such as an unexpected keyword, missing operator, or incorrect syntax. We observed a somewhat different pattern of outcomes in students resolving the CS1525 compiler error.

When students in the experimental condition dealt with this error while solving tasks with the availability of GPT hints (Phase 1), they outperformed the control group (Figure 4).

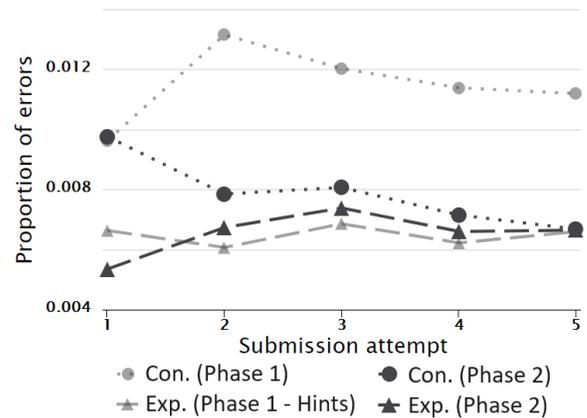

**Figure 4: Error CS1525 "Invalid expression term '{0}'". Proportion of Errors Across First 5 Submission Attempts (cumulative).**

In this case, as in the previous examples, the experimental group performed better than the control group for tasks with GPT hints disabled (Phase 2). However, the control group showed an improvement in performance across these problems, a pattern not seen for the experimental group.

*3.3.4 CS0266 "Cannot implicitly convert type '{0}' to '{1}'.* This error is encountered when there is an attempt to assign a value of one data type to a variable of another data type, and the compiler cannot perform an implicit conversion between them.

This error, unlike the much simpler syntax errors examined earlier, requires a deeper conceptual understanding of data types and their conversions in programming. For fixing the CS0266 compiler error, we observed varied results across different conditions.

When tasks were completed with the aid of GPT-generated hints (Phase 1), the experimental group, which had access to these hints, again performed slightly worse than the control group that did not receive support (Figure 5).

For the more difficult tasks where GPT hints were phased out (Phase 2), both groups demonstrated lower performance overall. In their first attempt at tasks with GPT hints disabled, the two groups showed similar levels of performance, indicating a baseline competency in addressing this complex error. In subsequent attempts, however, the control group displayed improvement in performance, while the experimental group unexpectedly showed a decrease.

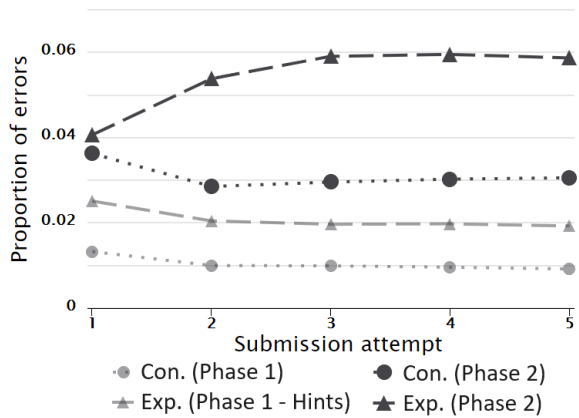

Figure 5: Error CS0266 "Cannot implicitly convert type '{0}' to '{1}'. An explicit conversion exists (are you missing a cast?)". Proportion of Errors Across First 5 Submission Attempts (cumulative).

*3.3.5 CS0161 "'{0}': not all code paths return a value".* This error is another example of a simpler syntax error occurring in a method that is expected to return a value, but where there is at least one code path that does not end with a *return* statement. Within the tasks where GPT hints were enabled for the experimental group (Phase 1), we observed the following pattern: Initially, in the first attempt, the experimental group slightly outperformed the control group. However, in subsequent attempts, the performance of the experimental group declined, while the control group maintained a consistent level of performance (Figure 6). When the GPT hints were disabled (Phase 2), we observed a similar pattern as for the first three errors introduced earlier: the performance of the experimental group showed a distinct improvement. The performance of the control group remained largely unchanged.

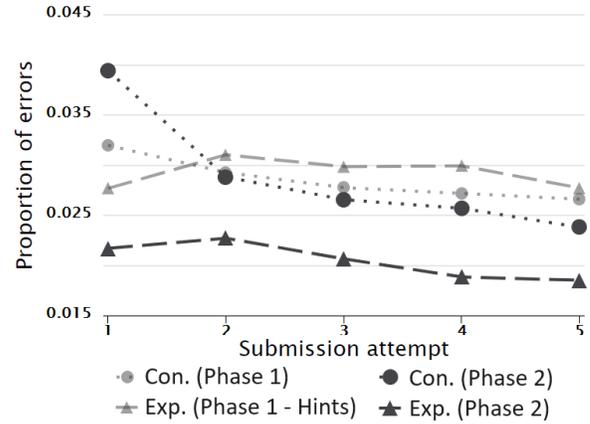

Figure 6: Error CS0161 "'{0}': not all code paths return a value". Proportion of Errors Across First 5 Submission Attempts (cumulative).

*3.3.6 CS1026 ") expected".* We also examined student responses to the CS1026 compiler error indicating that a closing parenthesis is missing – another example of a simpler syntax error.

For tasks where GPT-generated hints were available to the experimental group (Phase 1), we observed that the control group, which did not have access to these hints, exhibited a lower level of performance (Figure 7).

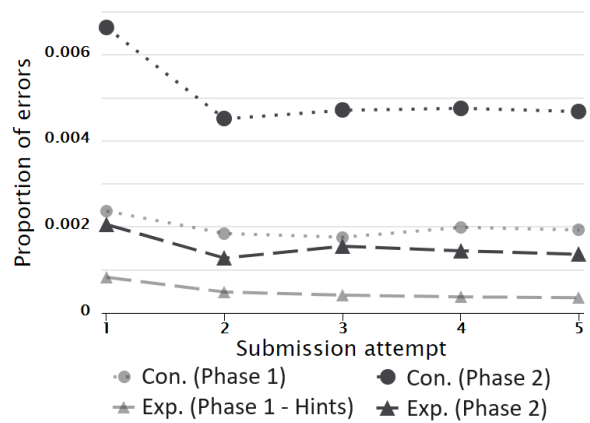

Figure 7: Error CS1026 ") expected". Proportion of Errors Across First 5 Submission Attempts (cumulative).

In a second set of tasks where GPT feedback was no longer provided for the experimental group (Phase 2), the experimental group again outperformed the control group. The performance level of the experimental group remained consistent with their earlier attempts when GPT hints were available. In contrast, the control group's performance deteriorated significantly.

### 3.4 Summary

In conclusion, this section analyzed six distinct errors and found that the experimental group exhibited enhanced performance in five instances when GPT-provided hints were withheld. This improvement predominantly occurred in errors related to syntax. However, addressing the CS0266 error, which necessitates a more profound comprehension of data types, did not follow this trend. It appears that the provided GPT assistance, may not have been helpful to students in learning to resolve more complex errors.

## 4  Discussion

Given the complexity of programming, including the large number of potential compiler errors, it is a challenge to provide concise and effective assistance to novices in the form of targeted hints for error resolution. Large language models have the potential to address this issue. In this study we presented the findings of an experiment that involved the use of GPT-4 hints generated in Polish aimed to assist novices within a platform for automated assessment of programming assignments and analyzed the hints' effectiveness at helping students resolve compiler errors.

According to a survey conducted to assess the perceived usefulness of GPT-generated hints, a significant majority of students positively rated the availability and utility of these hints, most generally considering them to be either *very* or *extremely* useful (56%). However, it is important to note that approximately 20% of the students generally did not find the hints beneficial. This suggests that while GPT hints are generally perceived as useful, there is a need for further improvement of their relevance and accuracy. Addressing this gap could involve strategies such as refined prompt engineering or fine-tuning of the AI model to better address diverse needs of students. This finding aligns with the outcomes reported in [18], underscoring the notion that while AI tools like GPT may positively stimulate the learning process, their efficacy is mixed and could be possibly augmented through more targeted improvements tailored to user feedback and specific learning contexts.

We also observed that the availability of GPT hints significantly impacted affective states of students directly after obtaining an error. We observed increased level of *focus* and also a significant decrease in *confrustion* (confusion and frustration) in the experimental group after receiving results for a submission with a compiler error. This outcome suggests that availability of the assistance provided by GPT hints positively influences the emotional state of learners. However, the relatively small sample of these responses suggests that further replication and investigation is warranted.

The results of the student performance analysis for different compiler errors were mixed, indicating that the availability of GPT hints did not consistently lead to improved performance across tasks with GPT hints enabled. Contrary to our expectations, we observed lower performance in the experimental condition for several error types. This outcome suggests that while GPT hints can be beneficial, their effectiveness in enhancing understanding and error resolution skills may be limited to specific types of compiler errors.

However, for more challenging tasks in the second half of each module where GPT hints were disabled, the performance of the experimental group was consistently higher for 5 out of 6 analyzed errors, with the exception of the "Error CS0266 'Cannot implicitly convert type '{0}' to '{1}'. An explicit conversion exists (are you missing a cast?)'." This error presents a different type of challenge than the remaining 5, simpler syntax-based errors. Resolving the CS0266 error, which involves the inability to implicitly convert between data types, requires a deeper conceptual understanding of data types and their conversions in programming. Our findings suggest that while our current GPT hints may help students learn to identify and rectify straightforward syntax or scope errors, their efficacy in fostering a comprehensive grasp of more complex concepts, such as those needed to resolve the CS0266 error, appears limited. Further work will be needed to enhance the quality and depth of the hints, as well as possibly adopting more integrated approaches that merge the benefits of AI tools with other teaching methods to cultivate a thorough understanding of more advanced programming concepts.

Our findings suggest that although the immediate and clear benefits of GPT-assisted learning are not universal, exposure to such hints can equip students with lasting skills or strategies that are beneficial when AI support is no longer available. The experimental group in our study showed improved error troubleshooting capabilities for syntax-based errors while solving more complex problems, unlike the control group.

There were several limitations of this study. First, the study's duration was brief, encompassing only three months of a four-month semester. Future research should consider capturing the longer-term effects of GPT-generated hints on student performance. The limited number of participants in this study, and its focus on a single university and course, also represents a limitation to the robustness and external validity of the findings. Future studies should therefore aim to involve a larger and more diverse number of participants. Finally, the focus of this study was exclusively on the C# programming language, while introductory computer science education frequently includes other languages such as Java, Python and C++. To provide a more comprehensive understanding of the general effectiveness of GPT-generated hints for programming education, future research should expand to include these additional programming languages and generating hints in other languages than Polish.

Overall, our study highlights the complexity and challenges related to the implementation of generative AI tools in computer science education, suggesting that their benefits might extend beyond direct assistance to fostering deeper learning and error resolution skills.

## ACKNOWLEDGMENTS

This paper was written with the assistance of ChatGPT, which was used to improve the writing clarity and grammar of first drafts written by humans. All outputs were reviewed and modified by two human authors prior to submission.